\documentclass[%
reprint,
superscriptaddress,
 amsmath,amssymb,
 aps,
]{revtex4-1}

\usepackage{graphicx}
\usepackage{dcolumn}
\usepackage{bm}
\usepackage{hyperref}
\usepackage{color}
\usepackage{amssymb}
\usepackage{amsmath}
\usepackage[T1]{fontenc}
\usepackage{bbold}
\usepackage{braket}
\DeclareMathOperator{\Tr}{Tr}

\begin{document}

\setlength{\textfloatsep}{1pt}

\newcommand{\Id}{\mathbb{1}}



\title{Fermionic Hamiltonians for quantum simulations: a general reduction scheme}

\author{Panagiotis Kl. Barkoutsos}
\affiliation{IBM Research - Zurich, S\"aumerstrasse 4, 8803 R\"uschlikon, Switzerland}
\affiliation{Institute for Theoretical Physics, ETH Zurich, 8093 Zurich, Switzerland}
 
\author{Nikolaj Moll}
\affiliation{IBM Research - Zurich, S\"aumerstrasse 4, 8803 R\"uschlikon, Switzerland}

\author{Peter W.J. Staar}
\affiliation{IBM Research - Zurich, S\"aumerstrasse 4, 8803 R\"uschlikon, Switzerland}

\author{Peter Mueller}
\affiliation{IBM Research - Zurich, S\"aumerstrasse 4, 8803 R\"uschlikon, Switzerland}

\author{Andreas Fuhrer}
\affiliation{IBM Research - Zurich, S\"aumerstrasse 4, 8803 R\"uschlikon, Switzerland}

\author{Stefan Filipp}
\affiliation{IBM Research - Zurich, S\"aumerstrasse 4, 8803 R\"uschlikon, Switzerland}

\author{Matthias Troyer}
\email{troyer@phys.ethz.ch}
\affiliation{Institute for Theoretical Physics, ETH Zurich, 8093 Zurich, Switzerland}

\author{Ivano Tavernelli}
\email{\{bpa, nim, taa, pmu, afu, sfi, ita\}@zurich.ibm.com}
\affiliation{IBM Research - Zurich, S\"aumerstrasse 4, 8803 R\"uschlikon, Switzerland}

\date{\today}

\begin{abstract} 

Many-body fermionic quantum calculations performed on analog quantum computers are restricted by the presence of $k$-local terms, which represent interactions among more than two qubits. 
These originate from the fermion-to-qubit mapping 
applied to the electronic Hamiltonians. 
Current solutions to this problem rely on perturbation theory in an enlarged Hilbert space. 
The main challenge associated with this technique is that it relies on coupling constants with very different magnitudes. 
This prevents its implementation
in currently available architectures.
In order to resolve this issue, we present an optimization scheme that unfolds the $k$-local terms into a linear combination of 2-local terms, while 
ensuring the conservation of all relevant physical properties of the original Hamiltonian, with 
several orders of magnitude smaller variation of the coupling constants.

\end{abstract}

\pacs{Valid PACS appear here}

\maketitle

Recent advances in the fields of quantum simulation and quantum computing are driven by the emergence of scalable 
quantum computing platforms based for instance on superconducting or ion trap qubits
~\cite{Barends2014b, Kelly2015b, Corcoles2015b, Riste2015b, Barends2015b}.
This stimulated interest for the simulation of the electronic structure problem using these quantum devices~\cite{AspuruGuzik2005, Lanyon2010, Whitfield2011, Kassal2011, Babbush2014, Wecker2015, Babbush2016, Romero2017b}. 
However, the quantum simulation of fermionic systems
with current quantum hardware faces the problem of encoding fermionic Hamiltonians in a qubit lattice that inherently follows bosonic statistics. To address this issue, several fermion-to-spin mappings have been developed~\cite{Jordan1928, Bravyi2000}.  
The most commonly used scheme is the Jordan-Wigner transformation, which
maps a local fermionic Hamiltonian to a local spin Hamiltonian suitable for applications in analog quantum simulations and quantum computing in general~\cite{Jordan1928}. 
Such transformations introduce many-body interaction terms (known as $k$-local), which cannot be implemented in the current analog quantum simulator architectures~\cite{Georgescu2014}. 
This is a non-trivial problem that was analyzed by \textit{Kempe et.\ al.}~\cite{Kempe2006} .
A solution to this problem consists in the reduction of the $k$-local terms into a linear combination of 2-local terms in an enlarged Hilbert space. 
The most commonly used reduction scheme is based on the ``Perturbative Hamiltonian Gadgets'' (PHGs) introduced by \textit{Kempe et.\ al.\ ~}\cite{Kempe2006} and further extended by other research groups~\cite{Bravyi2008, JordanFarhi2008, Seeley2012, Cao2015, Babbush2013, Moll2016b}.

In PHGs auxiliary qubits (ancillas) are introduced, which add to the set of physical qubits that describe the original Hamiltonian increasing the degrees of freedom of the system.
The spectrum of the enlarged Hilbert space ($\mathcal{H}_p \otimes \mathcal{H}_a$) consists of a physical branch ($\mathcal{H}_p$) associated with the ancillas ground state separated from all other higher lying branches (corresponding to the excited ancillas' space, ) by an energy gap $\Delta$,
ensuring the non-mixture between the physical and the unphysical quantum states (see SM).

This method has the advantage of being analytic, providing a way of controlling the accuracy of the results by means of a single parameter. 
The main drawback is that its accuracy can only be improved at the expense of spreading the coupling strengths of the 2-local terms over several orders of magnitude, making these techniques unsuited for practical implementations.

In view of this, we develop a rigorous numerical scheme that provides an accurate reduction of fermionic Hamiltonians to 2-local qubit Hamiltonians while keeping full control of the values of the coupling strengths. 
This may enable 
the simulation of fermionic systems in an analog quantum computer architecture with all coupling strengths situated in an achievable magnitude range.

The Hamiltonian describing a many-electron system in second quantization is:
\begin{equation}
\hat H_{F} = \hat H_F^{(1)} + \hat H_F^{(2)} = \sum_{ij} t_{ij} \hat a_i \hat a_j^{\dagger} + \sum_{ijkl} u_{ijkl} \hat a_i \hat a_j^{\dagger} \hat a_k \hat a_l^{\dagger}
\label{eq:Second_quant_Ham}
\end{equation}
where $\hat a_i^{\dagger}$ and $\hat a_i$ are the creation and annihilation operators of an electron in the orbital $i$ and $t_{ij}$ and $u_{ijkl}$ parametrize the one- and two- electron interaction terms, $\hat H_F^{(1)} \ {\rm{and}} \ \hat H_F^{(2)} $, respectively. 
For the first term of Eq.\ (\ref{eq:Second_quant_Ham}) the spin Hamiltonian obtained using the Jordan-Wigner transformation becomes
\begin{equation}
\begin{split}
\hat H_F^{(1)} &= \sum_{\braket{i,j}}^{i>j} t_{ij} (\Id^{\otimes i - 1}\otimes \hat \sigma^{-} \otimes \hat \sigma_z^{\otimes j-i-1} \otimes \hat \sigma^+ \hat \sigma_z \otimes \Id^{\otimes N- j - 1} )\\ &+ \sum_{\braket{i,j}}^{i<j}  t_{ij} (\hat \Id^{\otimes i - 1} \otimes \hat \sigma^{+} \otimes \hat \sigma_z^{\otimes j-i-1} \otimes \hat \sigma^- \hat \sigma_z \otimes \hat \Id^{\otimes N- j - 1})
\end{split}
\label{eq:scalable_JW}
\end{equation}
where $N$ is the total number of orbitals. 
A similar procedure also applies to the interaction term $\hat H^{(2)}_F$. 
The corresponding Hamiltonian in qubit space will acquire the form 
\begin{equation}
\hat H_ F =\sum_{i=1}^{M} c_i \left( \hat P_1^i \otimes \hat P_2^i \otimes \dots \otimes \hat P_N^i \right)
\label{eq:Operators_Ham}
\end{equation}
where  
$c_i$ is the coefficient of each $k$-local term, $M$ it the total number of terms and $N$ the total number of qubits. $\hat P_\alpha^i$ represents a Pauli operator, with $\hat P_\alpha^i \in \{\hat \sigma_x, \hat \sigma_y, \hat \sigma_z , \hat \Id \}$ and $\alpha \in \mathbb{N}$ labels the position in the array of qubits. 
To simplify the notation, we introduce the following contracted form of the tensor products (for $\alpha<\beta$) 
\begin{equation}
\hat P_\alpha^i \otimes P_\beta^j = \hat \Id^{\otimes \alpha -1} \otimes \hat P^i_{\alpha} \otimes \hat \Id^{\otimes \beta - \alpha -1} \otimes \hat P^j_{\beta} \otimes \hat \Id^{\otimes N- \beta} .
\label{eq:locality_term}
\end{equation}
The locality $k_i$ of every term $i$ is defined as the number of operators that are different from $ \hat \Id$. 

Although Hilbert space is not affected by the above transformations, the order of many-body interaction terms increases
to ${\rm{max}}\{k_i\}$ for $i\in \{1,\dots ,M\}$ in Eq.\ \eqref{eq:Operators_Ham}. The value of $k_i$ varies depending on the transformation scheme used and on how the qubits are ordered and connected as it is evident from Eq.~\eqref{eq:scalable_JW} \cite{Whitfield2016, Havlicek2017, Bravyi2017b}. For the Jordan-Wigner transformation the locality is $\mathcal{O}(N)$.

The \textit{`gadgetization'} process is a hierarchical procedure that reduces $k$-local terms to 2-local terms in several steps ($k$-local $\rightarrow$ ($k$-1)-local)~\cite{Bravyi2008, Oliveira2008}.
The main difficulties associated with this approach  are (\textit{i}) to address simultaneously a linear combination of $k$-local terms (cross-gadgets~\cite{Cao2015}) and 
(\textit{ii}) to cope with a large spread of magnitudes of coefficients in the 2-local gadget Hamiltonian. 
The latter occurs because the 2-local interaction strength between physical qubits and ancillas grows polynomially with $\Delta$~\cite{Babbush2014}.
It is known~\cite{Babbush2013, Babbush2014, Cao2015} that for the convergence of the gadget spectrum very large values (of the order of $10^6$ to $10^8$ atomic units, a.u.) of the spectral gap $\Delta$ are required. 
However, the implementation of this 
large range of coupling strengths is not achievable in current superconducting qubit circuits nor in any other known quantum computer architecture~\cite{Georgescu2014}.
This work will improve on both points (\textit{i}) and (\textit{ii}), making the gadgetization procedure a valid approach for the implementation of $k$-local Hamiltonians (with $ k> 2$) on currently available analog quantum computer hardware. 
The number of ancillas and the nature of the required coupling terms (in the space $\mathcal H_p \otimes \mathcal H_a$) is chosen according to 
\textit{Cao et.\ al.\ ~}\cite{Cao2015}
since it also allows cross-gadgetization, namely the simultaneous gadgetization of two or more $k$-local terms.

In order to achieve  \textit{magnitude-homogeneity} of the coupling strengths, we propose a new scheme that uses embedding of $\mathcal{H}_p$ into a 2-local Hamiltonian in a larger Hilbert space spanned by additional ancilla 
states and also uses a numerical optimization process for the derivation of the coupling strengths. 
This procedure will guarantee  full control over the error (of the physical energy spectrum and the reduced density matrix) 
while keeping
the coupling strengths within a suitable range for practical purposes. 

Recursive reduction schemes
have been previously discussed in literature~\cite{Oliveira2008, Cao2015, Leib2016}. 
Each of the iterations requires addition of ancilla qubits with the effect of extending the Hilbert space.
The last step in the reduction process therefore deals with a Hamiltonian of the form
\begin{equation}
\hat H_{3L} =  \sum_{i=1}^{M_3} c_i \left( \hat P_{\gamma_i(1)}^i \otimes \hat P_{\gamma_i(2)}^i \otimes \hat P_{\gamma_i(3)}^i \right)
\label{eq:3-local_Ham}
\end{equation}
where $M_3$ is the total number of 3-local terms and $\gamma_i$ maps the three operators to the corresponding qubits in the sequence (using the simplified notation in Eq.~\eqref{eq:locality_term}).
For this last iteration, we assume 
that the final number of qubits is N (among which we have already introduced $k_m-3$ ancillas per $k$-local term).

The reduction of the 3-local terms consists in calculating the set of coefficients $d_i$ in the extended space $\mathcal{H}_p \otimes \mathcal{H}_a$ with $N+M_3$ qubits 
\begin{equation}
\hat H_{2L} = \sum_{i=1}^{M_2} d_i \left( \hat P_{\zeta_i(1)}^i \otimes \hat P_{\zeta_i(2)}^i  \right)
\label{eq:2-local_Ham}
\end{equation}
where  $\zeta_i$ maps one operator into $\mathcal{H}_p$ and the other into  $\mathcal{H}_a$ for each index $i$.

In this work, instead of applying perturbation theory for the derivation of the coefficients $d_i$ in Eq.~\eqref{eq:2-local_Ham} we design an optimization scheme 
which maps the original $k$-local Hamiltonian into a 2-local Hamiltonian (Eq.~\eqref{eq:2-local_Ham}) while restricting the value of physical observables to a desired error threshold. 

The key improvement of this procedure is that it allows to map $\hat H_{3L}$ (in $\mathcal{H}_p$) into $\hat H_{2L}$ (in $\mathcal{H}_p \otimes \mathcal{H}_a$) with arbitrary precision while keeping all coefficients
of the Hamiltonian within the same 
order of magnitude.
In the PHG approach, the error in the spectrum decreases linearly with an exponential increase of $\Delta$ (and ultimately of the coefficients $d_i$). In contrast to this, our approach allows to reach the same accuracy solely by tuning the value of the coefficients without increasing their spread.

We first consider the selection of terms derived in~\cite{Cao2015}, which defines the maps $\zeta_i$ for a subset $M_2'$ of terms in Eq.~\eqref{eq:2-local_Ham}.
We then complete the list of 2-local terms by adding a set of additional $M_2^{''}$ terms 
selected according to their relevance. 
This is assessed for instance by evaluating their contribution to the cost function in an optimization with ($M_2'+1$) 2-local terms.
The inclusion of the additional $M_2^{''}$ terms offers the possibility to perform the optimization in a larger parameter space leading to more accurate solutions with a 
better suited distribution of the coefficients $d_i$.

The cost function $D$ is composed by the linear combination of three distances. 
The first measures the average difference of low energy part of the eigenspectrum in $\mathcal{H}_p$ ($i=1,\dots i_m$) as computed with $\hat H_{3L} $ and $\hat H_{2L}$, where $i_m$ is the number of eigenvectors that span $\mathcal{H}_p$. 
The second computes the average error for the density matrix, $n$, associated to the eigenspace of $\mathcal{H}_p$. 
In $\mathcal{H}_p \otimes \mathcal{H}_a$, the density matrix corresponding to each eigen-subspace is obtained through a partial trace over $\mathcal{H}_a$.
Finally, the last component ensures the existence of a sizable energy gap between the upper part of the  lower energy spectrum in $\mathcal{H}_p$ and the lower edge of the spectrum in $\mathcal{H}_a$.

In summary, the coefficients $d_i$ are obtained by minimizing the locality-reduction-optimizer (LRO) cost function
\begin{subequations}
\begin{gather}
D=\min_{\{d_i\}} \left[ \alpha_1 C_1 + \alpha_2 C_2 + \alpha_3 C_3 \right]   \\
C_1 = \sum_{i=0}^{i_{\rm{m}}} (\epsilon_{i}^{p} - \epsilon_{i}^{p \otimes a})^2 \label{second:cost_function_c1}\\
C_2 = \sum_{i=0}^{i_{\rm{m}}} ||n_{i}^{p} - {\Tr_a}n_{i}^{p \otimes a}|| \label{second:cost_function_c2} \\
C_3 = \ \left | {\min_{i\in\{1\dots 2^{n_a}\}}}(\epsilon_i^{a}) - {\max_{i\in\{1\dots 2^{n_p}\}}}(\epsilon_{i}^p) \right | \label{second:cost_function_c3} 
\end{gather}
\label{eq:optimizer}
\end{subequations}
where $p$ stands for $\mathcal{H}_p$ and $p \otimes a$ for $\mathcal{H}_p \otimes \mathcal{H}_a$, $\epsilon_i$ are the eigenvalues and $n_i=|\xi_i\rangle \langle \xi_i|$ is
the density matrix associated with the $i^{\text{th}}$ eigenvector $|\xi_i\rangle$.
The weights $\alpha_1$, $\alpha_2$ and $\alpha_3$ can be varied according to the corresponding importance of each criterion. 

\begin{figure}[t] 
 \includegraphics[width=1.02\columnwidth]{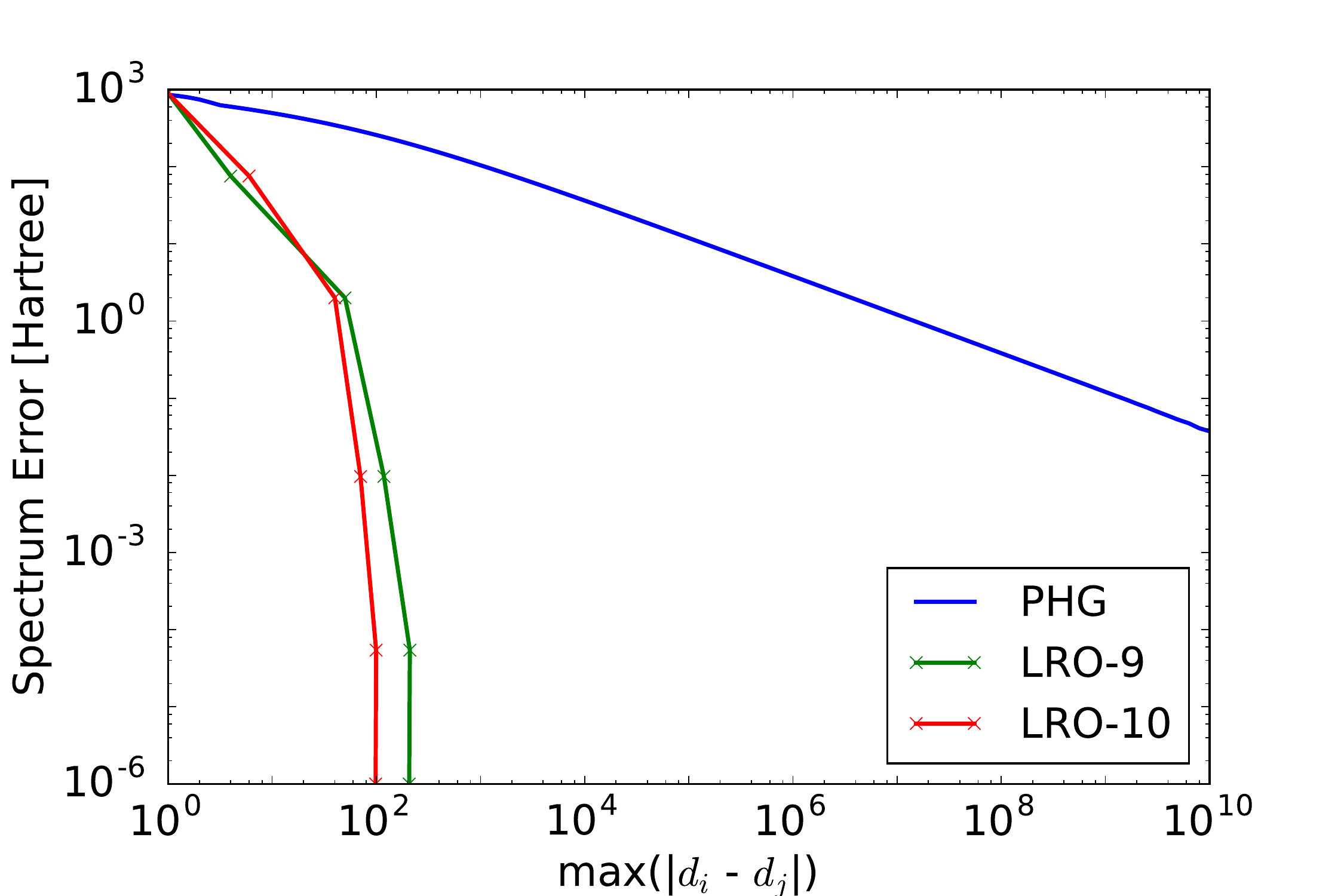}
\caption{
Total relative error in the energy spectrum $\left ( \sum_{i=1}^{i_m}  \left | \epsilon_i^{\text{LRO}} - \epsilon_i^{\text{exact}} \right |\right )$of the target Hamiltonian ($\hat{H}_{2L}$) as a function of the absolute maximum deviation $d_m$
of the optimized $d_i$ parameters, $d_m=\max_{ij}|d_i-d_j|$ for the PHG and LRO approaches. \label{fig:Total_error} }
\end{figure}

\begin{figure}[b] 
 \includegraphics[width=\columnwidth]{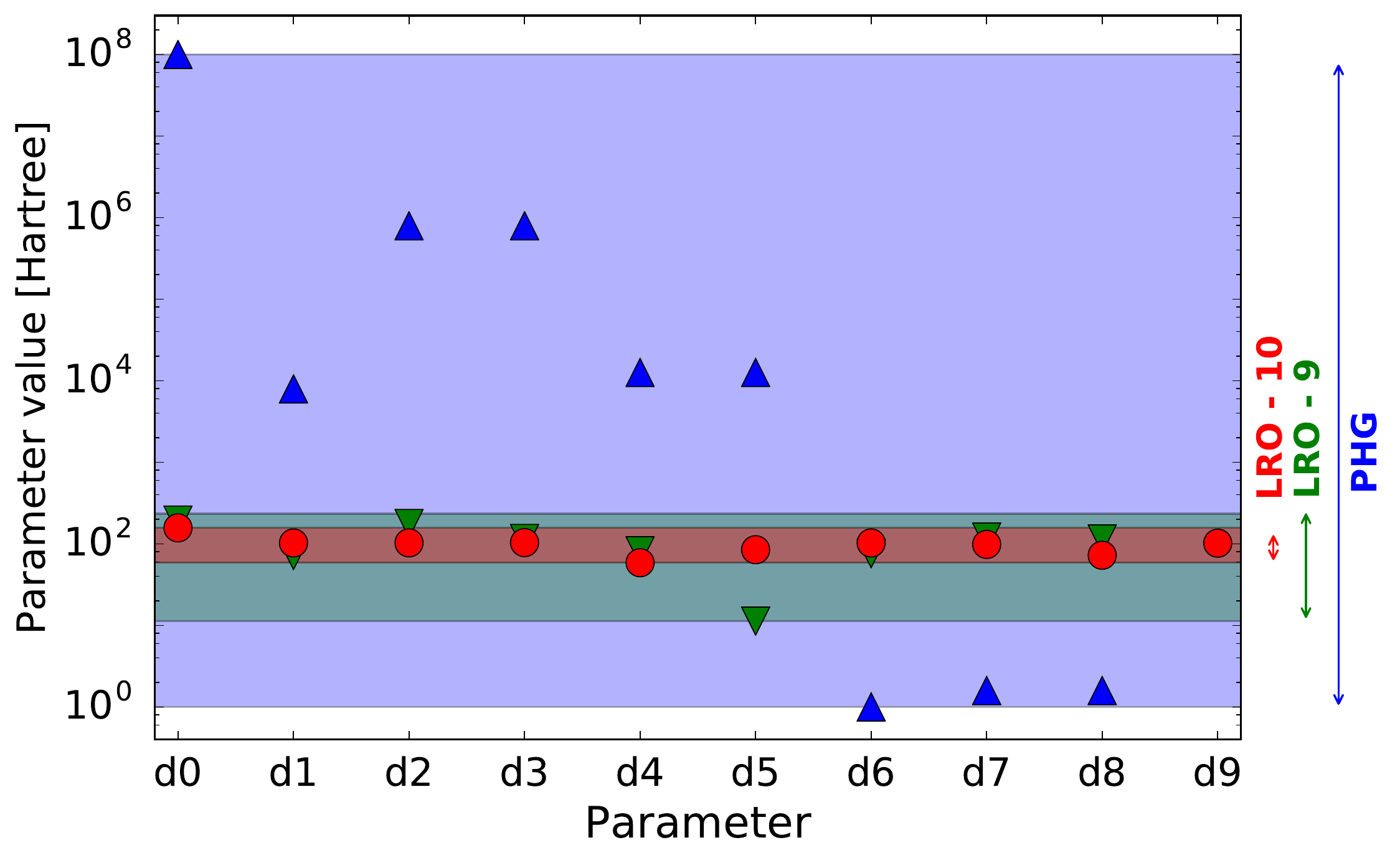}
\caption{Converged parameter sets $\{d_i\}_{i=0}^8$ for the 2-local terms in Eq.~\eqref{eq:2-local_Ham} corresponding to the $\hat H_{\text{target}}=\hat \sigma_{x(1)} \otimes \hat \sigma_{z(2)} \otimes \hat \sigma_{x(3)}$.
Color code: blue, PHG coefficients according to Cao~\cite{Cao2015}; green, LRO coefficients using 9 2-local terms; red, LRO coefficients using 10 2-local terms (that include the extra $\hat \sigma_{x(3)} \otimes \hat \sigma_{z(4)}$ term). 
The shaded areas highlight the different parameter ranges in log-scale. \label{fig:Parameter_values}}
\end{figure}

As a demonstration of the accuracy of our approach, we apply the optimization scheme to the reduction of a 3-local term of the form $\hat{H}_{3L}=\hat {\sigma_x}_{(1)} \otimes \hat {\sigma_z}_{(2)} \otimes \hat {\sigma_x}_{(3)}$, which typically arises in a fermionic Hamiltonian of a 3-qubit system (Eq. \eqref{eq:scalable_JW}). 
In a first test, we restrict the optimization to the same subspace of 2-local terms selected by PHG approach
~\cite{Cao2015}, which is spanned by 9 terms with coefficients $d_0$
to $d_8$ (LRO-9).

\begin{figure}[h]
\includegraphics[width=\columnwidth]{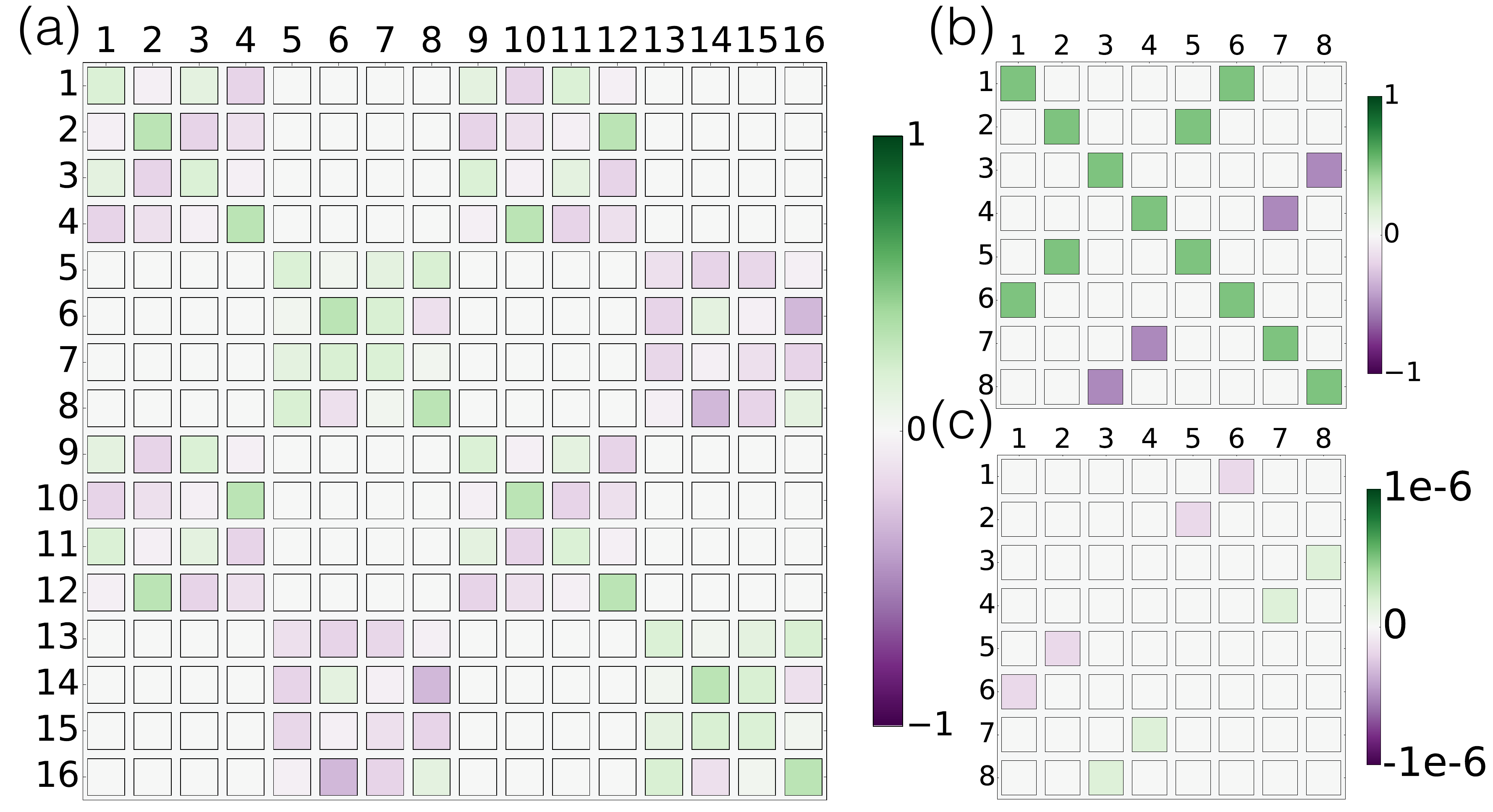}
\caption[width=0.92\columnwidth]{Analysis of the density matrix. (a) LRO converged density matrix $n_{(1-4)}$ of the full Hamiltonian in $\mathcal{H}_p \otimes \mathcal{H}_a $, and (b) corresponding reduced density matrix. The difference from the exact density matrix evaluated for reference Hamiltonian, $\hat H_{3L}$, is given in panel (c).}
\label{fig:Density_matrices}
\end{figure}

The optimization is performed using the cost function in Eq.~\eqref{eq:optimizer} with $i_m=8$  corresponding to the full spectrum of $\hat{H}_{\text{target}}$.
All criteria were considered equally important and the weight parameters ($\{\alpha_i\}_{i=1}^3$) were set to 1.
As starting conditions, we used uniform distributions of coefficients: $d'_i=1$, $d'_i=10^1$, $d'_i=10^2$, $d'_i=10^3$, and $d'_i=10^4$.
We observe that the cost function has several deep minima separated by large barriers that cannot be overcome by standard optimization schemes.
The value of the cost function at the different minima decreases uniformly as a function of the magnitude of the initial parameters and saturates for values of the initial conditions in the order of $d'_i=10^2$.
Fig.~\ref{fig:Total_error} shows the dependence of the relative error in lower part of the energy spectrum (first 8 eigenvalues) as a function of the absolute maximum deviation $d_m$
of the optimized $d_i$ parameters, $d_m=\max_{ij}|d_i-d_j|$.
The green dots correspond to the errors obtained at cost function values of $D=1, 10^{-2}, 10^{-4}, 10^{-6}$ and $10^{-8}$ a.u. (Eq.~\eqref{second:cost_function_c1}).
Interestingly, the LRO  approach can be pushed to very high accuracy ($< 10^{-6}$ Hartree) with $d_m$ values of the order of $10^2$.

The results can be further improved by adding an extra term of the form $ d_9 ( \hat \sigma_{x(3)} \otimes \hat \sigma_{z(4)} )$ in the optimization (LRO-10).
This term is selected as the one with the highest impact on each possible 2-local term of the total cost function. 
As expected, this additional degree of freedom allows for a further decrease of the cost function, while the minimum is obtained for a set of
parameters $\{d_i\}_{i=0}^{9}$ contained in a narrow range about $10^2$. 
Fig.~\ref{fig:Parameter_values} shows the spread of the parameters $d_i$ obtained using the different approaches. 
In general, LRO allows for a dramatic decrease of the spread of $d_m$, which goes from about $10^8$ for PHG to less than 40 for the LRO-9 approach.

To validate our results, we evaluate the density matrix $n_{(1-4)}$ associated with the 4-dimensional eigenspace corresponding to the lowest eigenvalue $\epsilon_1=-1$. 
Fig.~\ref{fig:Density_matrices} shows the LRO density matrix $n^{p \otimes a}_{(1-4)}$ in the $2^{(3+1)}$ dimensional Hilbert space $\mathcal{H}_p \otimes \mathcal{H}_a$ (Fig.~\ref{fig:Density_matrices}(a)) together with the corresponding reduced density matrix  ${n}^r_{(1-4)} = \Tr_a \sum_{i=1}^4 n_{i}^{p \otimes a}$ (Fig.~\ref{fig:Density_matrices}(b)). 
For the full converged solution, the LRO approach gives a maximum error $||n^p_{(1-4)}- {n}^r_{(1-4)}||$ in the density matrix of less than $10^{-6}$ (Fig.~\ref{fig:Density_matrices}(c)), where $n^p_{(1-4)}$ is the reference density matrix evaluated using the 3-local target Hamiltonian, $\hat{H}_{\text{target}}$.

\begin{figure}[t]
  \includegraphics[width=\linewidth]{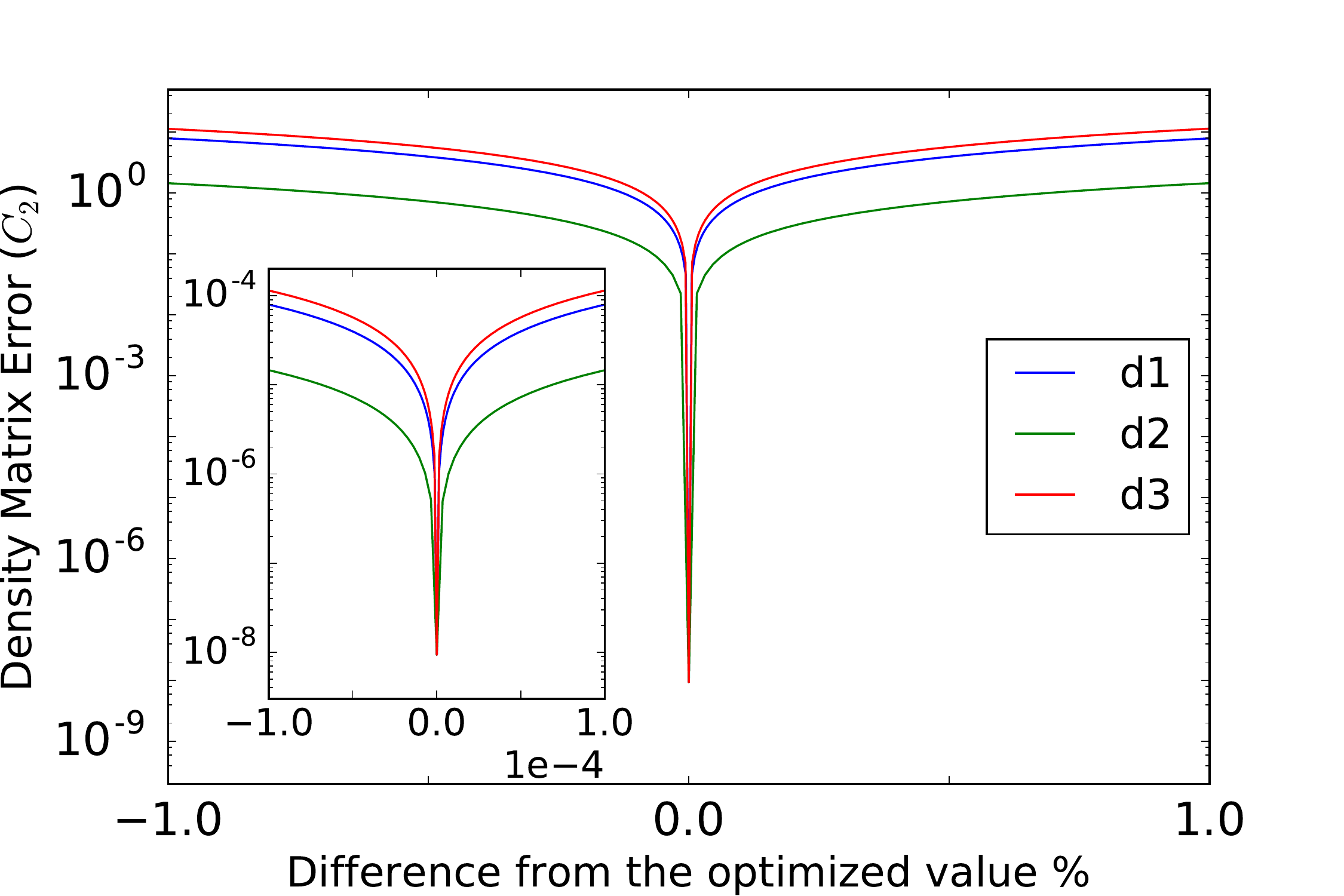}
\caption{\label{fig:Error_variation} Analysis of the dependence of error in the reduced density matrix $n^r_{(1-4)}$ on the 3 most sensitive parameters $d_1, d_2 \ \rm{and} \ d_3$.}
\end{figure}

Finally, we performed a stability analysis of the  LRO solution in order to assess the dependence of the accuracy in the physical observables (eigen-spectrum and density matrix) on the
numerical precision of the optimized parameters $d_i$.
This is of particular relevance if we think about the implementation of the transformed 2-local Hamiltonian into a physical quantum device. 
In this case, it is important to know what kind of precision for the coupling strengths is required in order to obtained a desired accuracy.
To this end, in Fig.~\ref{fig:Error_variation} we monitor the dependence of the total relative error of the reduced density matrix, $n_{(1-4)}$, as a function of the variation
in the parameters $d_1$ to $d_3$.
We observe that a variation of 1\% of the coefficients leads to about 10\% error in the density matrix. 
More interestingly, in order to achieve an accuracy of $10^{-6}$ for  ${n}_{(1-4)}$ a $10^{-4}\%$ precision in the coupling strengths, $d_i$, is required.
While this level of accuracy is easily achieved numerically, it is not yet clear if experimental implementations will be able to reach it as well.

In conclusion, the simulation of fermionic systems using a quantum computer requires the transformation of the original Hamiltonian into an equivalent spin Hamiltonian.
This procedure introduces $k$-local interaction terms among qubits, which pose a serious challenge to their implementation in available quantum computer  architectures.  
In this work, we derived a numerical approach for the reduction of $k$-local to 2-local terms in spin Hamiltonians. 
The scheme is applied iteratively, reducing the $k$-locality in steps of 1. 
At each iteration, the Hilbert space is enlarged through the addition of (at least) one additional ancilla qubit, which allows the reduction of the $k$-local terms into a linear combination of ($k-1$)-local terms.
The method is based on an optimization procedure (named LRO) that selects the most relevant ($k-1$)-local terms and optimizes the corresponding coefficients preserving the physical  properties (eigen-spectrum and density matrix) of the original system.
In particular, we show in full details how to perform the reduction of a 3-local term into a linear combination of 2-local term in the enlarged Hilbert space $\mathcal{H}_p \otimes \mathcal{H}_a$.
While the PHG method leads to a spread of coupling strengths $d_i$ that scales exponentially with the prescribed inverse error threshold, our method does not have this restriction.
In fact, our optimization approach leads to parameter sets that lie within the same order of magnitude and are fully compatible with the self-interaction terms ($1$-local) of the original physical Hamiltonian.
Finally, the analysis of the stability of the LRO solution reveals the level of accuracy required for the coefficients in order to be able to successfully simulate the transformed spin Hamiltonian on a realistic quantum computer. 

Although these strict requirements can possibly be a limiting factor for application in current analog quantum simulators, the use of the LRO scheme allows for a significant
reduction of the coefficients spread. This constitutes an important step towards future simulations of the electronic structure problems in analog quantum approaches.

\bibliographystyle{apsrev4-1}
%

\newpage

\

\newpage

\appendix

\section{Perturbative Hamiltonian Gadgets}
The reduction of $k$-local terms into series of $2$-local interactions requires embedding of the original Hamiltonian into an enlarged Hilbert space, $\mathcal{H}= \mathcal{H}_p \otimes \mathcal{H}_a$ span by the state vectors of the physical $n_p$-qubit Hamiltonian ($\in \mathcal{H}_p$) and  of the $n_a$-ancilla (auxiliary qubit) Hamiltonian ($\in \mathcal{H}_a$).
The PHG approach~\cite{Kempe2006, Bravyi2008, JordanFarhi2008,Cao2015, Oliveira2008, Leib2016} allows the generation of a gadget Hamiltonian ($\in \mathcal{H}$) of the form  $\tilde H= \hat H_0 + \hat V$ with $||\hat H_0 || \sim \mathcal{O}(\Delta) \gg ||\hat V|| \sim \mathcal{O}(1)$, 
such that the corresponding effective $k$-local Hamiltonian, $\hat H_{\text{eff}}$ obtained from perturbation expansion in $\hat V$, 
shares the same low energy spectrum as the original Hamiltonian $\tilde H$ in  $\mathcal{H}$. By matching the $k$-local $\hat H_{\text{eff}}$ Hamiltonian with the $k$-local physical Hamiltonian ($\hat H_{\text{eff}}\equiv \hat H_{\text{phys}} \otimes \hat \Pi$, where $\hat \Pi$ is the projector onto the ancilla ground space) it is possible to uniquely determine the structure of the 2-local gadget Hamiltonian $\tilde H$ in the enlarged Hilbert space $\mathcal{H}$. 
The spectral gap $\Delta$ controls the accuracy of the embedding of the $H_p$ into the gadget Hamiltonian $\tilde H$. 
$\hat H_0$ is proportional to $\Delta$ and therefore guarantees an energy separation between the physical and ancilla subspaces. 
Increasing $\Delta$,  the energy gap between the ground and excited ancilla spaces increases, improving the convergence of the perturbation expansion in $\hat V$ and therefore the match with the physical
eigenspectrum associated to the ancillas ground space.

Projecting the effective Hamiltonian, $\tilde H$ in the low energy subspace $\tilde H_-$ will reproduce the spectrum of the initial system, whereas the high energy subspace $\tilde H_+$ will be depended on the value of $\Delta$. 
\begin{equation}
\tilde{H}=
\begin{pmatrix}
\tilde{H}_{++} & \tilde{H}_{+-} \\
\tilde{H}_{-+} & \tilde{H}_{--}
\end{pmatrix}
H =
\begin{pmatrix}
H_{+} & 0\\
0  & H_{-}
\end{pmatrix}
V =
\begin{pmatrix}
V_{++} & V_{+-}\\
V_{-+}& V_{--}
\end{pmatrix}
\end{equation}

Here we employ the terms of the Hamiltonian as used by Cao et.\ al.\ \cite{Cao2015} for a $3$-body term of the form $\alpha (P_1 \otimes P_2 \otimes P_3)$ described as

\begin{equation}
 \hat H_0 = \Delta (\ket{1}\bra{1})_{u}
\label{eq:Penalty_ham}
\end{equation}

\begin{subequations}\label{first:main}
\begin{equation}
\begin{split}
 \hat V = \mu P_3 \otimes (\ket{1}\bra{1})_u+(\kappa P_1 + \lambda P_2) \otimes X_u+ \\+V_1 + V_2 + V_3
\end{split}
\tag{\ref{first:main}}
\label{eq:Perturbation_V}
\end{equation}
\begin{equation}
 \hat V_1 = \frac{1}{\Delta} (\kappa P_1 + \lambda P_2)^2  - \frac{1}{\Delta^2}(\kappa^2 + \lambda^2)\mu P_3 \label{first:Perturbation_V1}
\end{equation}
\begin{equation}
 \hat V_2 =-\frac{1}{\Delta^3} \left( \kappa P_1 + \lambda P_2 \right) ^4 \label{first:Perturbation_V2}
\end{equation}
\begin{equation}
 \hat V_3 = \sum^{m}_{i=1} \bar V_{ij}  \label{first:Perturbation_V3}
\end{equation}
\end{subequations}

where,

\begin{subequations}
\begin{equation}
\kappa = {\rm{sgn}}\left(a \right) \left ( \frac{|a|}{2} \right )^{\frac{1}{3}} \Delta^{\frac{3}{4}}
\end{equation}
\begin{equation}
\lambda = \left ( \frac{|a|}{2} \right )^{\frac{1}{3}} \Delta^{\frac{3}{4}}
\end{equation}
\begin{equation}
\mu = \left ( \frac{|a|}{2} \right )^{\frac{1}{3}} \Delta^{\frac{1}{2}}
\end{equation}
\end{subequations}

Instead of calculating coefficients $\kappa, \lambda \ \text{and} \ \mu$ and fixing the coefficients of every term of the Hamiltonian analytically, we numerically vary the coefficients of each of the above terms.

\end{document}